\documentclass[aps,pre,floats,preprint,showpacs,superscriptaddress]{revtex4}

\usepackage{epsfig}
\usepackage{latexsym,amsmath}

\begin{document}
\title{Modeling the Internet}

\author{M. \'Angeles Serrano}

\affiliation{School of Informatics, Indiana University,\\ Eigenmann
Hall, 1900 East Tenth Street, Bloomington, IN 47406, USA}

\author{Mari{\'a}n Bogu{\~n}{\'a}}

\affiliation{Departament de F{\'\i}sica Fonamental, Universitat de
  Barcelona,\\ Mart\'{\i} i Franqu\`es 1, 08028 Barcelona, Spain}

\author{Albert D\'{\i}az-Guilera}

\affiliation{Departament de F{\'\i}sica Fonamental, Universitat de
  Barcelona,\\ Mart\'{\i} i Franqu\`es 1, 08028 Barcelona, Spain}

\date{\today}

\begin{abstract}
We model the Internet as a network of interconnected Autonomous
Systems which self-organize under an absolute lack of centralized
control. Our aim is to capture how the Internet evolves by
reproducing the assembly that has led to its actual structure and,
to this end, we propose a growing weighted network model driven by
competition for resources and adaptation to maintain functionality
in a demand and supply ``equilibrium''. On the demand side, we
consider the environment, a pool of users which need to transfer
information and ask for service. On the supply side, ASs compete to
gain users, but to be able to provide service efficiently, they must
adapt their bandwidth as a function of their size. Hence, the
Internet is not modeled as an isolated system but the environment,
in the form of a pool of users, is also a fundamental part which
must be taken into account. ASs compete for users and big and small
come up, so that not all ASs are identical. New connections between
ASs are made or old ones are reinforced according to the adaptation
needs. Thus, the evolution of the Internet can not be fully
understood if just described as a technological isolated system. A
socio-economic perspective must also be considered.
\end{abstract}


\maketitle

\section{Introduction}
In an attempt to bring nearer theory and reality, many researchers
working on the new and rapidly evolving science of complex networks
\cite{Mendesbook} have recently shifted focus from unweighted graphs
to weighted networks. Commonly, interactions between elements in
network-representable complex real systems -may they be
communication systems, such as the Internet, or transportation
infrastructures, social communities, biological or biochemical
systems- are not of the same magnitude. It seems natural that the
first more simple representations, where edges between pairs of
vertices are quantified just as present or absent, give way to more
complex ones, where edges are no longer in binary states but may
stand for connections of different strength.

Weight is just one of the relevant ingredients in bringing network
modeling closer to reality. Others come from the fact that real
systems are not static but evolve. As broadly recognized, growth and
preferential attachment are also key issues at the core of a set of
recent network models focusing on evolution under an statistical
physics approach
\cite{Barabasi99,Huberman,Goh,Capocci,Fayed,Medina,Zhou,Yook}. This
models have been able to approximate some topological features
observed in many real networks --specifically the small-world
property or a power-law degree distribution-- as a result of the
organizing principles acting at each stage of the network formation
process. Although a great step forward in the understanding of the
laws that shape network evolution, these new degree driven models
cannot describe other empirical properties. Further on, in order to
achieve representations that closely match reality, it is necessary
to uncover new mechanisms.

Following this motivation, we believe that the general view of
networks as isolated systems, although possibly appropriate in some
cases, must be changed if we want to describe in a proper way
complex systems which not generate spontaneously but self-organize
within a medium in order to perform a function. Many networks evolve
in an environment to which they interact and which usually provides
the clues to understand functionality. Therefore, rules defined on
the basis of internal mechanisms alone, such as preferential
attachment that acts internally at the local scale to connect nodes
trough edges, are not enough. When analyzing the dynamics of network
assembly, the interlock of its constituents with the environment
cannot be systematically obviated.

With the aim of approaching applicability, in this work we blend all
ideas above to present a growing network model in which both, nodes
and links, are weighted \cite{Serrano}. The dynamical evolution is
driven by exponential growth, competition for resources and
adaptation to maintain functionality in a demand and supply
``equilibrium'', key mechanisms which may be relevant in a wide
range of self-organizing systems, in particular those where
functionality is tied to communication or traffic. The medium in
which the network grows and to with it interacts is here represented
by a pool of elements which, at the same time, provide resources to
the constituents of the network and demand functionality, say for
instance users in the case of the Internet \cite{Romusbook} or
passengers in the case of the world-wide airport network
\cite{Barrat04a}. Competition is here understood as a struggle
between network nodes for new resources and is modeled as a rich get
richer (preferential attachment) process. For their part, this
captured elements demand functionality so that nodes must adapt in
order to perform efficiently. This adaptation translates into the
creation of weighted links between nodes.

In this work, we apply those ideas to the Internet. In the realm of
complexity theory, the Internet is a paradigmatic example and
significant efforts has been devoted to the development of models
which reproduce the topological properties observed in its maps
\cite{Romusbook}. Candidates run from topology generators
\cite{Waxman,Inet} to degree driven growing networks models
\cite{Fitness,GBA} or Highly Optimized Tolerance (HOT) models
\cite{HOT}. Some of them reproduce heavy-tailed degree distributions
and small-world properties, but perform poorly when estimating
correlations or other characteristic properties, such as the
rich-club phenomenon or the k-core structure. By contrast, we will
show that our model nicely reproduces an overwhelming number of
observed topological features: the small-world property, the
scale-free degree distribution $P(k)$, high clustering coefficient
$c_k$ that shows a hierarchical structure, disassortative
degree-degree correlations, quantified by means of the average
nearest neighbors degree of nodes of degree $k$, $\bar{k}_{nn}(k)$
\cite{Alexei}, the scaling of the higher order loop structure
recently analyzed in \cite{Bianconi}, the distributions of the
betweenness centrality, $P(b)$, and triangles passing through a
node, $P(T)$, and, finally, the k-core decomposition uncovering its
hierarchical organization \cite{kcore,Lanet}.

We will consider the Internet evolution at the Autonomous System
(AS) level. ASs are defined as independently administered domains
which autonomously determine internal communications and routing
policies \cite{Romusbook} and, as a first approximation, we can
assign each AS to an Internet Service Provider (ISP). This level of
description means a coarse grained representation of the Internet.
Nevertheless, further detail is not necessary when aiming to explain
and predict the large-scale behavior. Thus, the network will be made
up of ASs as nodes connected among them with links which can be of
different strength or bandwidth. On the side of the environment
modeling, we place hosts on the level as users.

In the next sections we analyze the growth of the Internet over the
last years. Then we present the model. Working in the continuum
approximation, we find analytically the distribution of the sizes
(in number of users) of ASs and the degree distribution. Then, we
introduce an algorithm in order to simulate network assembly. At
this stage, we also make a first attempt to the consideration of
geographical constraints. Finally, the synthetic networks are
compared to the real maps of the Internet through a variety of
different measures.

\section{The growth of the Internet}
\begin{figure}[h]
\epsfig{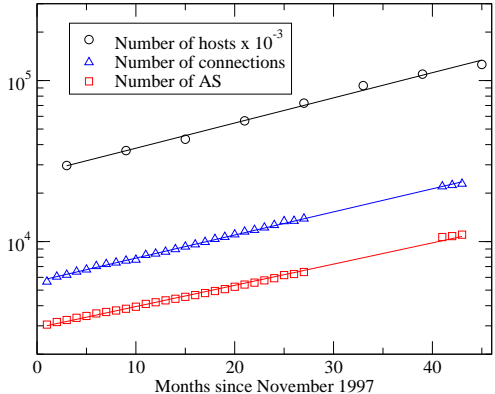} \caption{Temporal
evolution of the number of hosts, autonomous systems and connections
among them from November 1997 to May 2002. Solid lines are the best
fit estimates. Each point for the number of ASs and connections is
an average over one month of daily measurements. Error bars are of
the order of the symbol size.}
 \label{evolution}
\end{figure}

Let $W(t)$ be the total number of users in the environmental pool at
a given time $t$, measured as hosts. $N(t)$ and $E(t)$ stand for the
number of ASs and edges among them in the network, respectively.
Empirical measures for the growth in the number of users have been
obtained from the Hobbes' Internet Timeline \cite{Hosts}. The growth
of the network is analyzed from AS maps collected by the {\it Oregon
route-views} project, which has recorded data since November 1997
\cite{Evolution}. According to those observations, shown in
Fig.\ref{evolution}, we will assume exponential growths for these
quantities, $W(t)\approx W_0 e^{\alpha t}$, $N(t)\approx N_0
e^{\beta t}$, and $E(t)\approx E_0 e^{\delta t}$. These exponential
growths, in turn, determine the scaling relations with the system
size: $W \propto N^{\alpha/\beta}$, $E \propto N^{\delta/\beta}$ and
$\langle k \rangle \propto N^{\delta/\beta-1}$.

The rates of growth can be measured to be $\alpha=0.036\pm0.001$,
$\beta=0.0304\pm0.0003$, and $\delta=0.0330\pm0.0002$ (units are
month$^{-1}$), where $\alpha \gtrsim \delta \gtrsim \beta$. These
three rates are quite close to each other but they are not equal. In
fact, the inequality $\alpha \gtrsim \beta$ must hold in order to
preserve network functionality. When the number of users increases
at a rate $\alpha$, there are two mechanisms capable to compensate
the demand they represent: the creation of new nodes and the
creation of new connections by nodes already present in the network.
When both mechanisms take place simultaneously, the rate of growth
of new nodes, $\beta$, as well as the rate for the number of
connections, $\delta$, must necessarily be smaller than $\alpha$.
Any other situation would lead to an imbalance between demand and
supply of service in the system. On the other hand, in a connected
network, $\delta$ must be equal or greater than $\beta$. If $\delta$
equals $\beta$ the average number of connections per node, or
average degree, remains constant in time, whereas it increases when
$\delta \gtrsim \beta$. This increase could correspond to a demand
per user which is not constant but grows in time, probably due to
the increase of the power of computers over time and, as a
consequence, to the ability to transfer bigger files or to use more
demanding applications.

\section{The model}
We define our model according to the following rules: (i) At rate
$\alpha W(t)$, new users join the system and choose node $i$
according to some preference function, $\Pi_i(\{\omega_j(t)\})$,
where $\omega_j(t)$, $j=1,\cdots,N(t)$, is the number of users
already connected to node $j$ at time $t$. The function
$\Pi_i(\{\omega_j(t)\})$ is normalized so that $\sum_i
\Pi_i(\{\omega_j(t)\})=1$ at any time. (ii) At rate $\beta N(t)$,
new nodes join the network with an initial number of users,
$\omega_0$, randomly withdrawn from the pool of users already
attached to existing nodes. Therefore, $\omega_0$ can be understood
as the minimum number of users required to keep nodes in business.
(iii) At rate $\lambda$, each user changes his AS and chooses a new
one using the same preference function $\Pi_i(\{\omega_j(t)\})$.
Finally, (iv) each node tries to adapt its number of connections to
other nodes according to its present number of users or size, in an
attempt to provide them an adequate functionality.

With all specifications above, we will work in the continuum
approximation to find some analytic results, specifically the
distribution of the sizes of ASs and the degree distribution.

\subsection{Analytic results}
The resource dynamics of single nodes is described by the following
stochastic differential equation
\begin{equation}
\frac{d
\omega_i}{dt}=A(\omega_i,t)+\left[D(\omega_i,t)\right]^{1/2}
\xi(t), \label{langevin}
\end{equation}
where $\omega_i$ is the number of users attached to AS $i$ at time
$t$. The time dependent drift is $A(\omega_i,t)=(\alpha+\lambda)
W(t) \Pi_i-\lambda \omega_i-\beta \omega_0$, and the diffusion term
is $D(\omega_i,t)=(\alpha+\lambda) W(t) \Pi_i+\lambda \omega_i+\beta
\omega_0-2\lambda \omega_i \Pi_i$. Application of the Central Limit
Theorem guaranties the convergence of the noise $\xi(t)$ to a
gaussian white noise in the limit $W(t) \gg 1$. The first term in
the expression for the drift is a creation term accounting for new
and old users that choose node $i$. The second term represent those
users who decide to change their node and, finally, the last term
corresponds to the decrease of users due to introduction of newly
created nodes. To proceed further, we need to specify the preference
function $\Pi_i(\{\omega_j(t)\})$. We assume that, as a result of a
competition process, nodes bigger in resources get users more easily
than small ones. The simplest function satisfying this condition
corresponds to the linear preference, that is,
$\Pi_i(\{\omega_j(t)\})=\omega_i/W(t)$, where $W(t)=\omega_0 N_0
\exp{(\alpha t)}$.  In this case, the stochastic differential
equation (\ref{langevin}) reads
\begin{equation}
\frac{d \omega_i}{dt}=\alpha \omega_i-\beta
\omega_0+\left[(\alpha+2\lambda)\omega_i+\beta \omega_0
\right]^{1/2} \xi(t). \label{langevin2}
\end{equation}
Notice that reallocation of users ({\it i.e.} the $\lambda$-term)
only increases the diffusive part in Eq.~(\ref{langevin2}) but has
no net effect in the drift term, which is, eventually, the leading
term. The complete solution of this problem requires to solve the
Fokker-Planck equation corresponding to Eq.~(\ref{langevin2}) with a
reflecting boundary condition at $\omega=\omega_0$ and initial
conditions $p(\omega_i,t_i |
\omega_0,t_i)=\delta(\omega_i-\omega_0)$ ($\delta(\cdot)$ stands for
the Dirac delta function). Here $p(\omega_i,t | \omega_0,t_i)$ is
the probability that node $i$ has a number of users $\omega_i$ at
time $t$ given that it had $\omega_0$ at time $t_i$. The choice of a
reflecting boundary condition at $\omega=\omega_0$ is equivalent to
assume that $\beta$ is the overall growth rate of the number of
nodes, that is, the composition of the birth and dead processes
ruling the evolution of the number of nodes.

Finding the solution for this problem is not an easy task.
Fortunately, we can take advantage of the fact that, when $\alpha
> \beta$, the average number of users of each node increases exponentially
and, since $D(\omega_i,t) = {\cal O} \left( A(\omega_i,t)\right)$,
fluctuations vanishes in the long time limit. Under this zero noise
approximation, the number of users connected to a node introduced at
time $t_i$ is
\begin{equation}
\omega_i(t|t_i)=\frac{\beta}{\alpha}\omega_0+(1-\frac{\beta}{\alpha})\omega_0
e^{\alpha(t-t_i)}.
\end{equation}
The probability density function of $\omega$ can be calculated in
the long time limit as
\begin{equation}
p(\omega,t)=\beta e^{-\beta t} \int_0^t e^{\beta t_i}
\delta(\omega-\omega_i(t|t_i)) dt_i
\end{equation}
which leads to
\begin{equation}
p(\omega,t)= \displaystyle{ \frac{\tau(1-\tau)^{\tau}
\omega_0^{\tau}}{(\omega-\tau\omega_0)^{1+\tau}}
\Theta(\omega_c(t)-\omega)}, \label{p_omega}
\end{equation}
where we have defined $\tau\equiv \beta/\alpha$ and the cut-off is
given by $\omega_c(t) \sim (1-\tau)\omega_0 e^{\alpha t} \sim W(t)$.
Thus, in the long time limit, $p(\omega,t)$ approaches a stationary
distribution with an increasing cut-off that scales linearly with
the total number of users. The exponent $\tau$ depends on the
relative values of $\beta$ and $\alpha$, which can be different but
typically would stay close so that $\tau$ would value around 2.

The key point now is to construct a bridge between the competition
and the adaptation mechanisms, in other words, to see how to relate
the number of users attached to an AS with its degree and bandwidth.
Our basic assumption is that vertices are continuously adapting
their strength or bandwidth, the total weight of its connections, to
the number of users they have. However, once a node decides to
increase its bandwidth it has to find a peer who, at the same time,
wants to increase its bandwidth as well. The reason is that
connection costs among nodes must be assumed by both peers. This
fact differs from other growing models in which vertices do not ask
target vertices if they really want to form those connections. Our
model is, then, to be thought of as a coupling between a competition
process for users and adaptation of vertices to their current
situation, with the constraint that connections are only formed
between ``active'' nodes, that is, those ASs with a positive
increment of their number of users. Let $b_i(t|t_i)$ be the total
bandwidth of a node at time $t$ given that it was introduced at time
$t_i$. This quantity can include single connections with other
nodes, {\it i. e.} the topological degree $k$, but it also accounts
for connections which have higher capacity. This is equivalent to
say that the network is, in fact, weighted and $b_i$ is the weighted
degree. To simplify the model we consider that bandwidth is
discretized in such a way that single connections with high capacity
are equivalent to multiple connections between the same nodes. Then,
when a pair of nodes agrees to increase their mutual connectivity
the connection is newly formed if they were not previously connected
or, if they were, their mutual bandwidth increases by one unit,
reinforcing in this way their connectivity. Now, we assume that, at
time $t$, each node adapts its total bandwidth proportionally to its
number of users, or size, following a lineal relation. Thus, we can
write
\begin{equation}
b_i(t|t_i)=1+a(t)\left( \omega_i(t|t_i)-\omega_0 \right).
\label{bandwidth}
\end{equation}
Summing Eq.~(\ref{bandwidth}) for all nodes we get $a(t)
=(2B(t)-N(t))/(W(t)-\omega_0N(t)) \approx 2B(t)/W(t)$, where $B(t)$
is the total bandwidth of the network. $B(t)$ is, obviously, an
upper bound to the total number of edges of the network. This
suggests that $B(t)$ will grow according to $B(t)=B_0 e^{\delta'
t}$. As the number of users grows, the global traffic of the
Internet also grows, which means that nodes do not only adapt their
bandwidth to their number of users but to the global traffic of the
network. Therefore, $a(t)$ must be an increasing function of $t$,
which, in turn, implies that $\delta'>\alpha$ and, thus,
$\delta'>\delta$. As a consequence, the network must necessarily
contain multiple connections. This can be explicitly seen by
inspecting the scaling of the maximum bandwidth, which reads
$b_c(t)\propto N(t)^{\delta'/\beta}$, that is, faster than $N(t)$.
Therefore, the topological degree of a node cannot be proportional
to its bandwidth. Nevertheless, it is clear that $k_i$ and $b_i$ are
positive correlated random variables. We then propose that degree
and bandwidth are related, in a statistical sense, through the
following scaling relation
\begin{equation}
k(t|t_i) = \left[b(t|t_i) \right]^{\mu}, \mbox{ \hspace{0.2cm}
$\mu<1$, } \label{degree_wealth}
\end{equation}
which implies that all nodes can form multiple connections,
regardless of their size. This scaling behavior has recently been
observed in other weighted networks \cite{Barrat04a,Barrat04b}. The
superlinear behavior of  $b_c(t)$, combined with this scaling
relation, ensures that rich ASs will connect to a macroscopic
portion of the system, so that the maximum degree will scale
linearly with the system size. Empirical measurements made in
\cite{Goh} showed such linear scaling in the AS with the largest
degree. This sets the scaling exponent to $\mu=\beta/\delta'$.

All four growth rates in the model are not independent but can be
related by exploring the interplay between bandwidth, connectivity,
and traffic of the network. Summing Eq.~(\ref{degree_wealth}) for
all vertices, the scaling of the total number of connections is
$E(t)\propto N(t)^{2-\alpha/\delta'}$, which leads to
$\delta'=\alpha \beta /(2\beta-\delta)$. Combining this relation
with Eqs.~(\ref{p_omega}), (\ref{bandwidth}) and
(\ref{degree_wealth}), the degree distribution reads
\begin{equation}
P(k)\approx \frac{\tau(1-\tau)^{\tau} \left[ \omega_0
a(t)\right]^{\tau}}{\mu} \frac{1}{k^{\gamma}} \Theta(k_c(t)-k)
\label{p_k}
\end{equation}
for $k \gg 1$, where the exponent $\gamma$ takes the value
$\gamma=1+1/(2-\delta/\beta)$. Strikingly, the exponent $\gamma$ has
lost any direct dependence on $\alpha$ becoming a function of the
ratio $\delta/\beta$. Using the empirical values for $\beta$ and
$\delta$, the predicted exponent is $\gamma =2.2 \pm 0.1$, in
excellent agreement with the values reported in the literature
\cite{Falou99,Alexei}. Of course, this does not mean that the
exponent $\gamma$ is independent of $\alpha$, since both, $\beta$
and $\delta$, may depend on the growth of the number of users.
Anyway, our model turns out to depend on just two independent
parameters which can be expressed as ratios of the rates of growth,
$\beta/\alpha$ and $\delta/\beta$.

\subsection{Simulations}
So far, we have been mainly interested in the degree distribution of
the AS map but not in the specific way in which the network is
formed. To fill this gap we have performed numerical simulations
that generate network topologies in nice agreement with real
measures of the Internet. Although ASs are distributed systems, we
assume they follow the same spatial distribution as the one measured
for routers, so that we are able to define a physical distance among
them to take into account connection costs \cite{Yook}. Our
algorithm, following the lines of the model, works in four steps:
\begin{enumerate}
\item At iteration t, $\Delta W(t)=\omega_0 N_0 (e^{\alpha
t}-e^{\alpha(t-1)})$ users join the network and choose provider
among the existing nodes using the linear preference rule. \item
$\Delta N(t)=N_0(e^{\beta t}-e^{\beta(t-1)})$ new ASs are introduced
with $\omega_0$ users each, those being randomly withdrawn from
already existing ASs. Newly created ASs are located in a two
dimensional plane following a fractal set of dimension $D_f=1.5$
\cite{Yook}. \item Each AS evaluate its increase of bandwidth,
$\Delta b_i(t|t_i)$, according to Eq.~(\ref{bandwidth}). \item A
pair of nodes, $(i,j)$, is chosen with probability proportional to
$\Delta b_i(t|t_i)$ and $\Delta b_j(t|t_j)$ respectively, and,
whenever they both need to increase their bandwidth, they form a
connection with probability $D(d_{ij},\omega_i,\omega_j)$. This
function takes into consideration that, due to connection costs,
physical links over long distances are unlikely to be created by
small peers. Once the first connection has been formed, they create
a new connection with probability $r$, whenever they still need to
increase their bandwidth. This step is repeated until all nodes have
the desired bandwidth.
\end{enumerate}
It is important to stress the fact that nodes must be chosen with
probability proportional to their increase in bandwidth at each
step. The reason is that those nodes that need a high bandwidth
increase will be more active when looking for partners to whom form
connections. Another important point is the role of the parameter
$r$. This parameter takes into account the balance between the costs
of forming connections with new peers and the need for
diversification in the number of partners. The effect of $r$ in the
network topology is to tune the average degree and the clustering
coefficient by modulating the number of multiple connections. The
exponent $\gamma$ is unaffected except in the limiting case $r
\rightarrow 1$. In this situation, big peers will create a huge
amount of multiple connections among them, reducing, thus, the
maximum degree of the network. Finally, we chose an exponential form
for the distance probability function $D(d_{ij},\omega_i,\omega_j)
=e ^ {-d_{ij}/d_c(\omega_i,\omega_j)}$, where
$d_c(\omega_i,\omega_j)= \omega_i\omega_j/\kappa W(t)$ and $\kappa$
is a cost function of number of users per unit distance, depending
on the maximum distance of the fractal set. All simulations are
performed using $\omega_0=5000$, $N_0=2$, $B_0=1$, $\alpha=0.035$,
$\beta=0.03$, and $\delta'=0.04$. The final size of the networks is
$N \approx 11000$, which approximately corresponds to the size of
the actual maps for 2001 that we are considering in this work.

\section{Testing the model}
To test the model we construct synthetic networks from our algorithm
with and without taking into consideration the geographical
distribution of ASs, and we contrast several measures on those
graphs to those of real maps, more specifically, the AS map dated
May 2001 from data collected by the Oregon Route Views Project
\cite{Evolution}, and the AS extended (AS+) map \cite{Qian} which
completes the previous one with data from other sources. Let us note
that all the measures presented here are performed over the same
synthetic networks. The parameters of the model are fixed once and
for all before generating the networks so that they are not tuned in
order to approach different properties.

\begin{figure}[t]
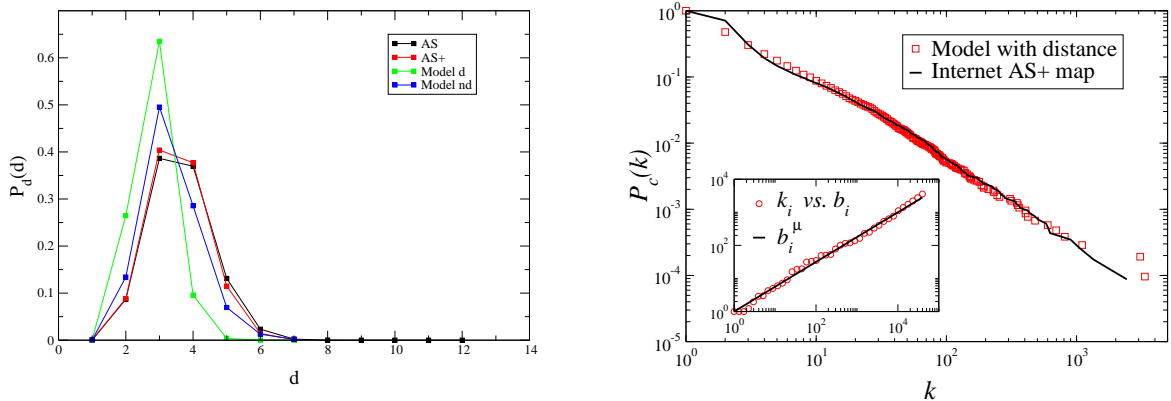

\begin{minipage}[t]{3.2in}
\vspace*{-5.2cm} \epsfig{file=SerranoFig2.eps, width=5cm, angle=-90}
\end{minipage}
\hfill
\begin{minipage}[t]{3.2in}
\epsfig{file=SerranoFig3.eps, width=7.2cm}
\end{minipage}
\vspace*{-1cm}\caption{Distribution of the shortest path lengths
(left) and cumulative degree distribution ($P_c(k)=\sum_{k'\geq k}
P(k')$) (right) for the extended AS map compared to simulations of
the model, $r=0.8$. Inset (right): Simulation results of the AS's
degree as a function of AS's bandwidth. The solid line stands for
the scaling relation Eq.~(\ref{degree_wealth}) with
$\mu=\beta/\delta'=0.75$.}
 \label{p_d_p_c}
\end{figure}

\begin{figure}[h]
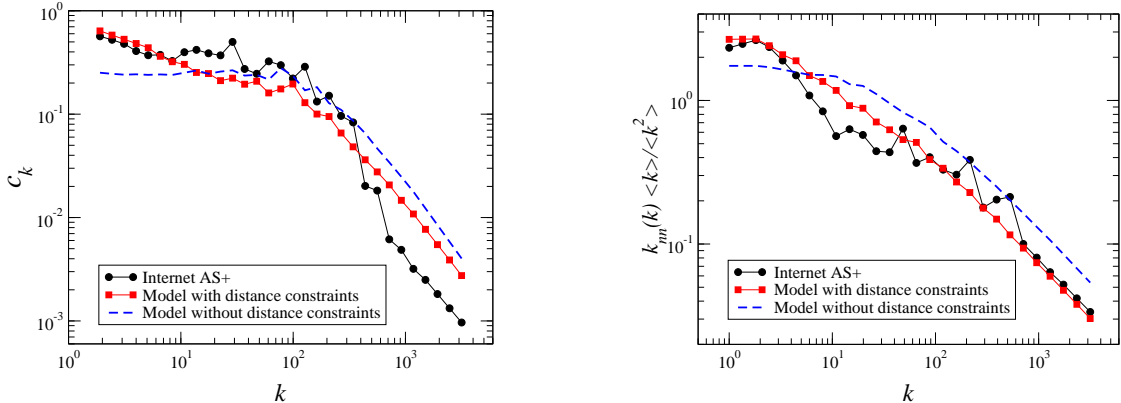

\begin{minipage}[t]{3.2in}
\epsfig{file=SerranoFig4.eps, width=6.5cm}
\end{minipage}
\hfill
\begin{minipage}[t]{3.2in}
\epsfig{file=SerranoFig5.eps, width=6.5cm}
\end{minipage}
\vspace*{-1cm} \caption{Clustering coefficient, $c_k$, (left), and
normalized average nearest neighbors degree, $\bar{k}_{nn}(k)
\langle k \rangle / \langle k^2 \rangle$, (right), as functions of
the node's degree for the extended autonomous system map (circles)
and for the model with and without distance constraints (red squares
and dashed line, respectively).} \label{c_k-knn}
\end{figure}

First, we analyze a first category of measures which include the
features of traditional interest when aiming to reproduce the
Internet topology. The small world effect becomes clear when
analyzing the distribution of the shortest path lengths, as seen in
the left side graph of Fig.~\ref{p_d_p_c}, with an average shortest
path length very close to the real one. The graph on the right of
Fig.~\ref{p_d_p_c} shows simulation results for the cumulative
degree distribution, in nice agreement to that measured for the AS+
map. The inset exhibits simulation results of the AS's degree as a
function of the AS's bandwidth, confirming the scaling {\it ansatz}
Eq.~(\ref{degree_wealth}). Clustering coefficient and average
nearest neighbors degree are showed in Fig.~\ref{c_k-knn}. Dashed
lines result from the model without distance constraints, whereas
squares correspond to the model with distance constraints.
Interestingly, the high level of clustering coming out from the
model arises as a consequence of the pattern followed to attach
nodes, so that only those AS willing for new connections will link.
As can be observed in the figures, distance constraints introduce a
disassortative component by inhibiting connections between small ASs
so that the hierarchical structure of the real network is better
reproduced.

\begin{figure}
\begin{minipage}[t]{3.2in}
\epsfig{file=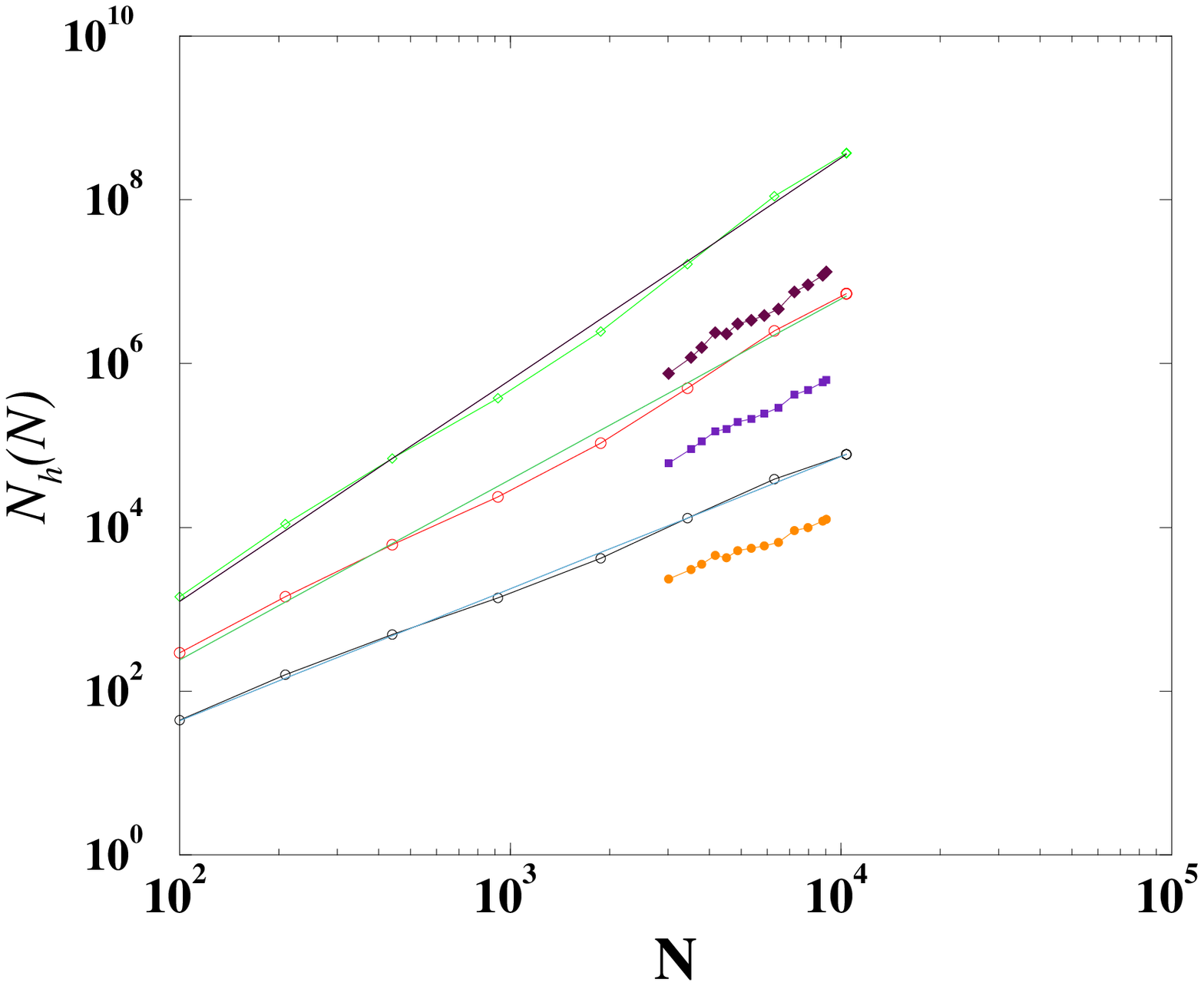, width=6.5cm}
\end{minipage}
\hfill
\begin{minipage}[t]{3.2in}
\epsfig{file=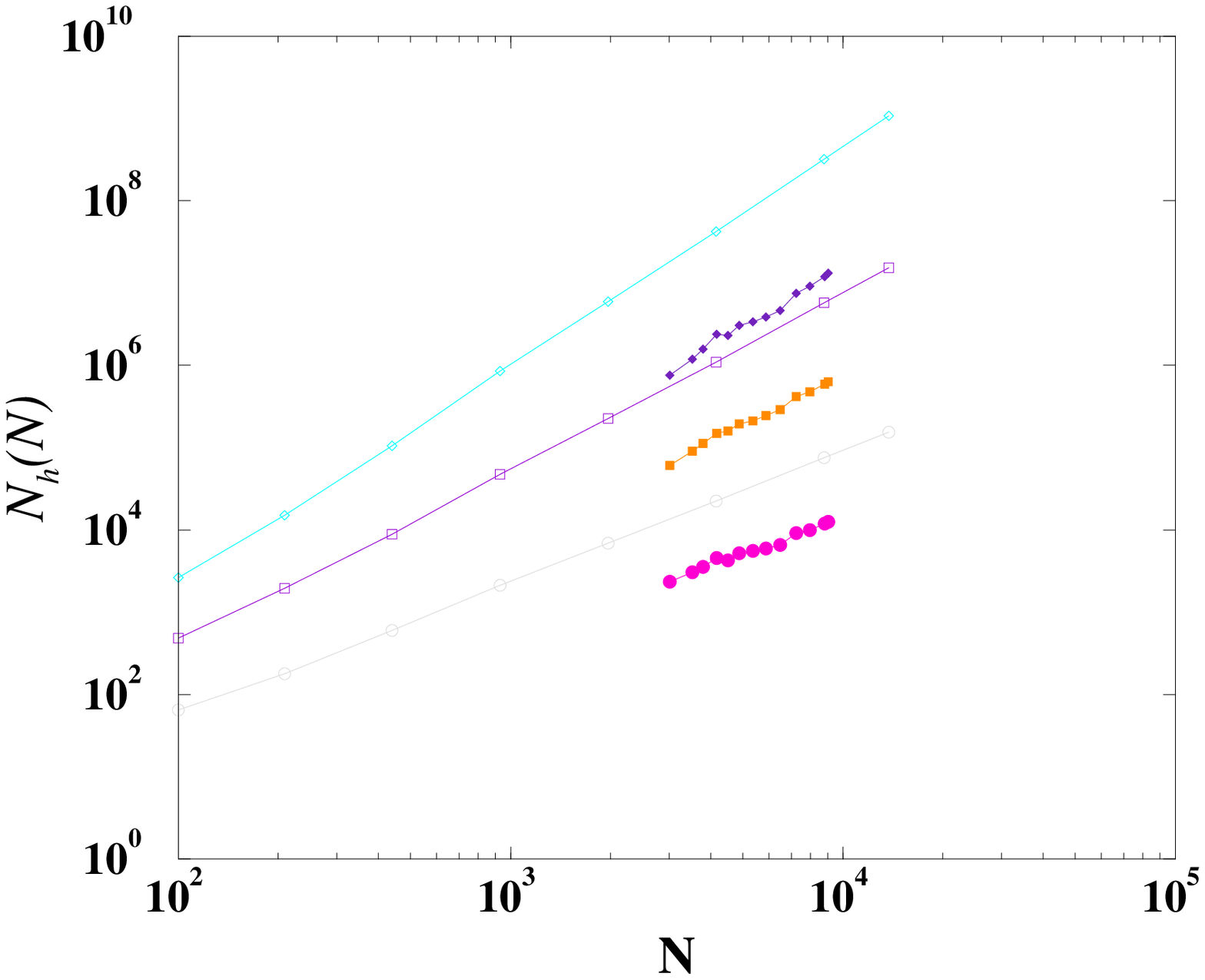, width=6.5cm}
\end{minipage}
\vspace*{-1.2cm}
 \caption{Scaling of the number of loops of size 3,
4 and 5 for the model with and without distance constraints, on the
left and on the right respectively. Short lines correspond to real
measures.}
 \label{nh}
\end{figure}

\begin{figure}[ht]
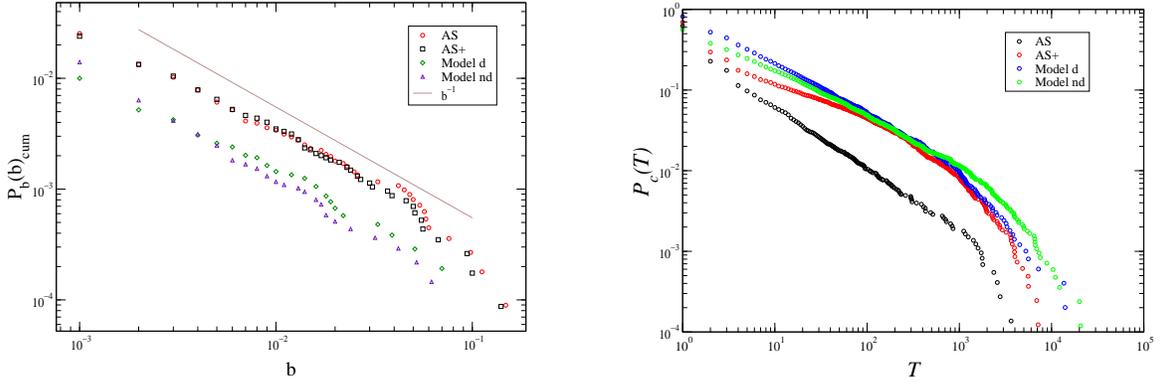

\begin{minipage}[t]{3.2in}
\epsfig{file=SerranoFig8.eps, width=5cm, angle=-90}
 \end{minipage}
\hfill
\begin{minipage}[t]{3.2in}
\epsfig{file=SerranoFig9.eps, width=5cm, angle=-90}
\end{minipage}
\caption{Cumulative distributions of the betweenness centrality
(left) and of the number of triangles passing by a node (right).}
\label{pb_ptri}
\end{figure}

Now, we turn our attention to new measures, which run from the
scaling of higher orders loops to the k-core structure. Not only
two-point correlations are well approximated by our model, but it is
also able to reproduce the scaling behavior of the number of loops
of size 3, 4 and 5. This has been recently measured for the Internet
at the AS level in \cite{Bianconi}, and it is seen to follow a power
of the system size of the form $N_h(N)\sim N^{\xi(h)}$, with
exponents that are closely reproduced by our synthetic networks, see
Fig.~\ref{nh} and table \ref{hexponents}.
\begin{table}
\caption{Values for the exponents $\xi(h)$ for $h=3$, $4$, and $5$
for the Internet and the models with and without distance
constraints (after Bianconi {\it et al.} \cite{Bianconi}).}
\begin{tabular}{lccc}
  \hline \hline
  System & $\xi(3)$ & $\xi(4)$ & $\xi(5)$ \\ \hline
  Internet AS map& $1.45\pm 0.07$ & $2.07\pm 0.01$ & $2.45\pm 0.08$ \\
  Model with distance & $1.60\pm 0.01$ & $2.20\pm 0.03$ & $2.70\pm 0.03$ \\
  Model without distance& $1.59 \pm 0.03$ & $2.11\pm 0.03$ & $2.64 \pm 0.03$ \\
  \hline \hline
  \end{tabular}
\label{hexponents}
\end{table}
In Fig.~\ref{pb_ptri}, we observe on the left the cumulative
distribution of betweenness centrality as proposed by Freeman
\cite{Freeman}, a measure of the varying importance of the vertices
in a network. On the right, the cumulative distribution of triangles
passing by a node (for a discussion of the relevance of P(T) see,
for instance, \cite{Vazquez}).

\begin{figure}[h]
\begin{minipage}[t]{2in}
\epsfig{file=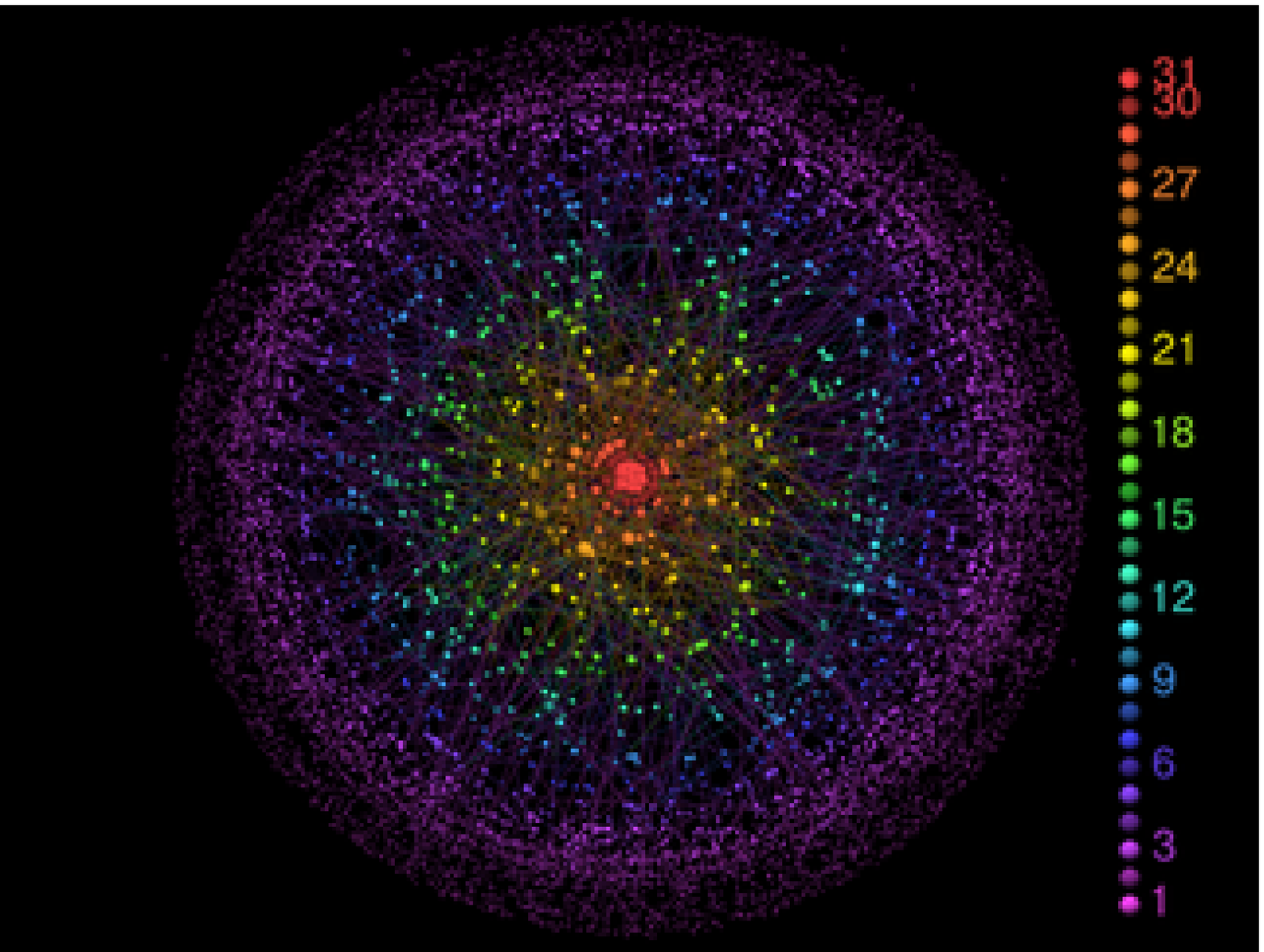, width=5.5cm}
\end{minipage}
\hfill
\begin{minipage}[t]{2in}
\epsfig{file=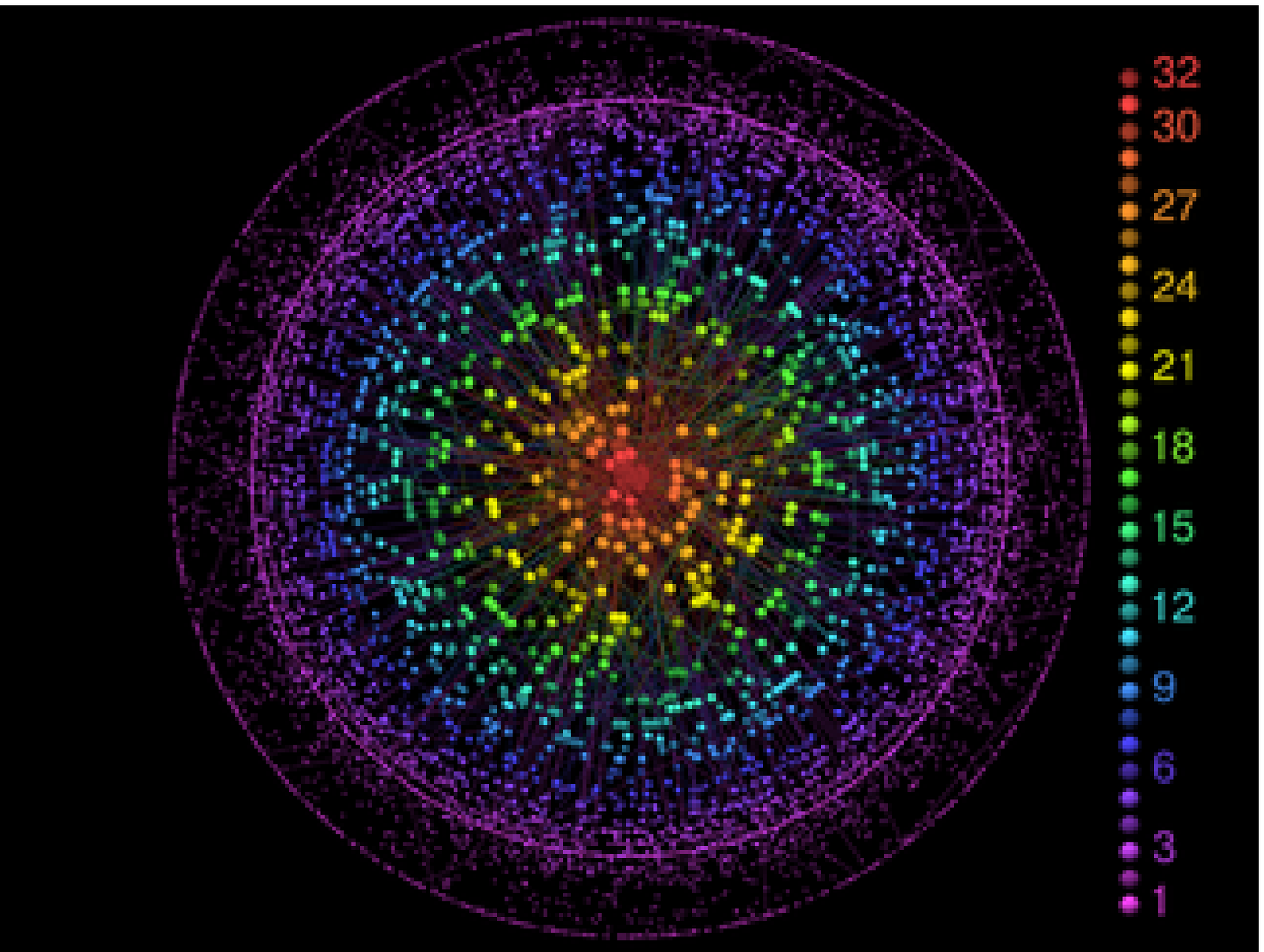, width=5.5cm}
\end{minipage}
\hfill
\begin{minipage}[t]{2in}
\epsfig{file=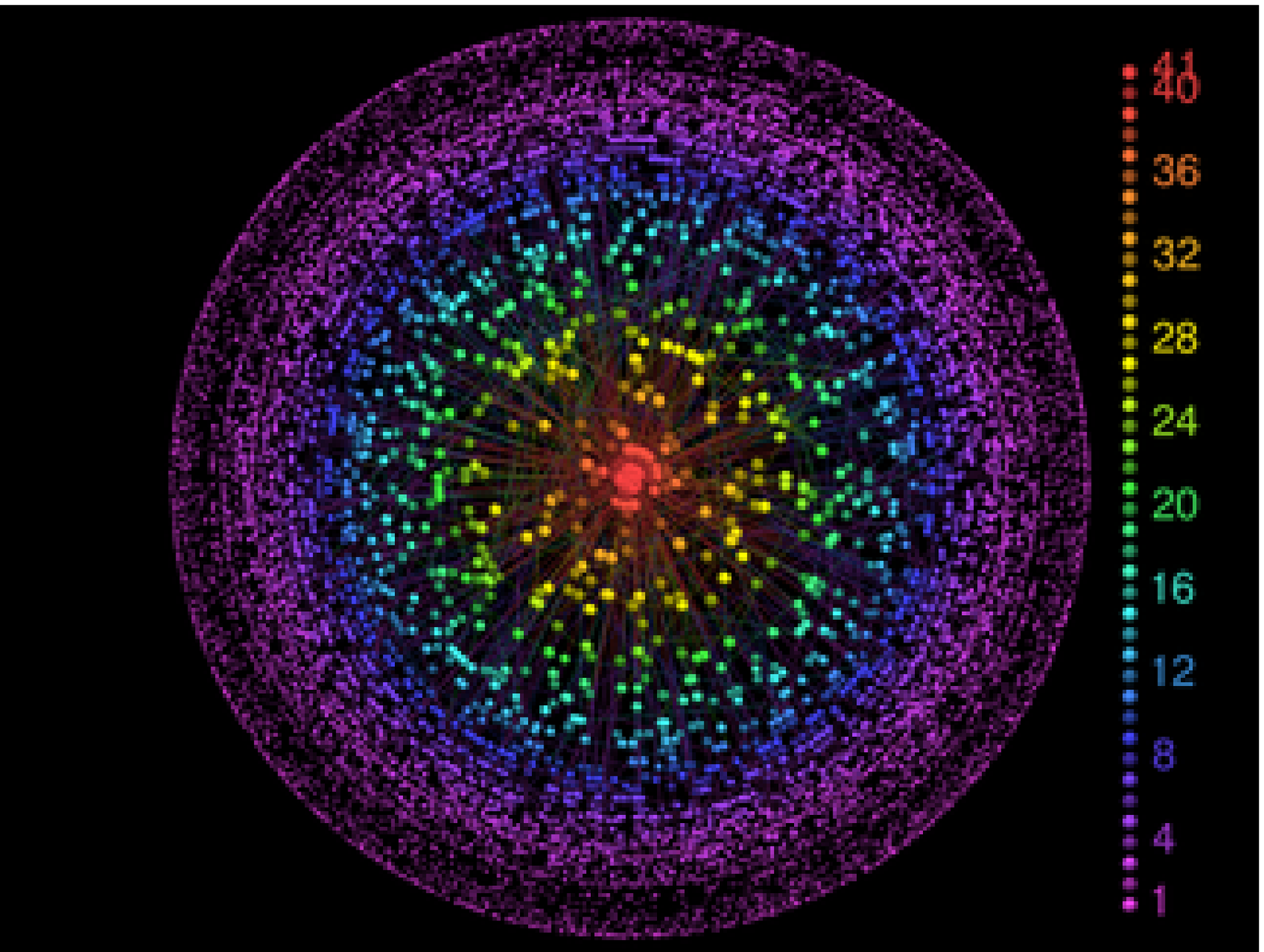, width=5.5cm}
\end{minipage}
 \caption{ k-core decompositions for the AS extended map of the Internet (left) and
for the maps generated from our model with and without distance
(center and right respectively). These visualizations have been
produced with the tool LANET-VI \cite{Lanet}.}
 \label{kcore}
\end{figure}

Finally, we also show the k-core decomposition of the actual and the
synthetic maps. The k-core decomposition is a recursive reduction of
the network as a function of the degree, which allows the
recognition of hierarchical structure and more central nodes
\cite{kcore}. A very good agreement between real measures and our
models can be appreciated in Fig.\ref{kcore}. In the case of the
model with distance constraints, even the coreness, the maximum
number of layers in the $k$-core decomposition, is almost the same
as in the Internet map. These visualizations have been produced with
the tool LANET-VI \cite{Lanet}.

\section{Conclusions}
In summary, we have presented a simple weighted growing network
model for the Internet, based on evolution, environmental
interaction and heterogeneity. The dynamics is driven by two key
mechanisms, competition and adaptation, which may be relevant in
other self-organizing systems. Beyond technical details, many
empirical features are nicely reproduced but open questions remain,
perhaps the more important one being whether the general ideas and
mechanisms exposed in this work could help us to better understand
other complex systems.

\begin{acknowledgments}
We acknowledge G. Bianconi for kindly provide us with
Figs.~\ref{nh}, and Ignacio Alvarez-Hamelin and Alessandro
Vespignani for valuable comments on the k-core decomposition. This
work has been partially supported by DGES of the Spanish government,
Grant No. FIS2004-05923-CO2-02 and Grant No. BFM-2003-08258, and
EC-FET Open project COSIN IST-2001-33555. M. B. acknowledges
financial support from the MCyT (Spain).
\end{acknowledgments}

\end{document}